\documentstyle[sprocl,epsfig]{article}
\def\L{\Lambda}

\def\beq{\begin{equation}}
\def\eeq{\end{equation}}
\def\beqra{\begin{eqnarray}}
\def\eeqra{\end{eqnarray}}
\begin{document}
\title{Gauge-Invariant Renormalization Group at Finite 
Temperature} 
\author{Massimo Pietroni}
\address{CERN-Theory Division, CH-1211 Geneva 23, Switzerland 
\footnote{Address 
after June, 1, 1997: Dipartimento di Fisica, Universit\`a di Padova, Via
 F. Marzolo 8, I-35131 Padova, Italy.}}    
\maketitle\abstracts{I 
describe an application of Wilson Renormalization group to
the real time formalism of finite temperature field theory. The approach 
has two nice features: 1) the RG flow equations describe 
non-perturbatively the effect of thermal fluctuations only, and,
2) the flow is gauge invariant. I then describe the application of 
the method to the study of the gluon 
self-energy in SU(N), and present results for the computation of 
the Debye and magnetic  screening masses.}

\section{Thermal coarse graining}
In this talk, I will describe an application of Wilson Renormalization
Group (RG) to high temperature field theory. The main purpose is  to construct
a tool for the non-perturbative resummation of  thermal fluctuations at 
larger and larger length scales applicable also to gauge 
theories~\cite{noi1,noi2}. In the case of non abelian gauge theories, 
the ultimate goal is to go below the 
scale at which the  Hard Thermal Loop (HTL) effective field theory 
breaks down.

Our approach is the following. We assume 
the quantum field theory in the vacuum is known, that is, we
have some approximation scheme ({\it e.g.} perturbation theory, 
lattice computations, ...)
that we trust in order to do computations and relate the physical observables
at $T=0$ to the parameters of the theory. We are interested 
in the purely thermal effects arising when the theory is put in a
thermal bath.
Moreover, in many physical applications, we are mainly 
interested in  long 
wavelength thermal fluctuations, since the short wavelength  ones
($\lambda$ much 
smaller than $1/T$) can be treated perturbatively.
Then, in the RG phylosophy, we will integrate out all the `irrelevant' 
degrees of freedom which, in this context, are given by all the quantum 
fluctuations and the thermal fluctuations of short wavelength. We will
 call this
procedure `thermal coarse graining'.

The appropriate framework for this program is  the real time formalism of high 
temperature field theory
(see ref.~\cite{RT} for a torough discussion). Consider 
the free propagator of a scalar particle,
\beq
D^{11}(k) = \frac{i}{k^2-m^2+i\varepsilon} + 2 \pi N(|k_0|) \delta(k^2-m^2) ,
\label{freeprop}
\eeq
(where I have written down the 1-1 component only). It has two important
features, which will play a crucial role in the following: 1) there is a clear
 separation between the `vacuum' contribution, and the 
`thermal' one, which contains the Bose-Einstein distribution function 
$ N(|k_0|)$, and, 2) the `thermal' part involves on-shell particles 
only.

Proceeding in close analogy to what is  done
in the application of the RG to $T=0$ quantum field theory~\cite{Pol},
 we modify the free propagator by introducing a cut-off function. The
peculiarity of this approach is that the cut-off acts  
{\it on the thermal part only} of the propagator, 
{\it i.e.}
\beq
{\bf D}(k) = {\bf D^0}(k) + {\bf D^T}(k) \; \longrightarrow \;
{\bf D_\L}(k) = {\bf D^0}(k) + \theta(|\vec{k}|,\L) {\bf D^T}(k)\,,
\eeq
where $ \theta(|\vec{k}|,\L)$ tends to $1$ for $\L \ll |\vec{k}|$ 
and to $0$ for $\L \gg |\vec{k}|$.
Inserting the modified propagator in the path integral expression for
the generating functional, we obtain a $\L$-dependent 
generating functional and, by the usual manipulations, a $\L$-dependent 
effective action.
%\beq
%Z_\L[{\rm J}] = \int {\cal D} \phi \exp\left(\frac{1}{2} 
%{\rm tr} \, \phi \cdot {\bf D_\L}^{-1} \cdot 
% \phi + i S[\phi] + i {\rm tr} \,  {\rm J} \cdot \phi \right).
%\eeq
These objects reduce to their counterparts 
for the $T=0$ {\it renormalized} quantum 
field theory in the 
$\L \rightarrow \infty$ limit, and to the ones of the quantum field theory in 
thermal  equlibrium at the temperature $T$ in the $\L \rightarrow 0$ limit.

The interpolation between the two limits is described by RG equations which, in
the case of one particle irreducible vertices can be obtained by the 
following simple recipe:
1) take the expression for the one loop correction to the desired vertex;
2) substitute the tree level propagators and vertices in it with the full, 
cut-off dependent ones;
3) take the derivative with respect to the explicit cut-off dependence 
({\it i.e.} derive only the cut-off function in the propagators).
To be more explicit, the RG equation for the tadpole in the scalar theory is 
given by
\beq
\L\frac{\partial \:\:\:\:}{\partial \L} \Gamma^{(1)}_\L = -\frac{1}{2}\int 
\frac{d^4 k}{(2 \pi)^4}\, \delta(|\vec{k}|-\L) \, N(|k_0|) \rho_\L(k) 
\,\epsilon(k_0)
\,\Gamma^{(3)}_\L(k, -k, 0)\,,
\label{flowtad}
\eeq
where we recognise the Bose-Einstein distribution function, the
full three-point function, and the spectral function $ 
\rho_\L(k)$, defined as the discontonuity of the full propagator across 
the real axis.

The physical meaning of the flow equation (\ref{flowtad}) is evident: the new
thermal modes coming into thermal equilibrium at the scale $\L$ are 
weighted by the full spectral function, induced
by all the quantum and thermal modes already integrated out.
 
\section{Gauge invariance}
A strong limitiation to the applicability of RG methods to gauge theories  
comes from the fact that the introduction of a momentum cut-off generally 
leads
to a breaking of BRS invariance. More precisely, new 
$\L$-dependent contributions to the Slavnov-Taylor (ST) 
identities appear, 
which vanish only in the limit in which the cut-off is removed and the
full theory is recovered. This is actually the situation in the 
 applications of the RG both at $T=0$ and at $T\neq 0$  in the 
imaginary time formalism~\cite{BRS}. In these cases, both on-shell and
off-shell modes are cut-off and this unavoidably spoils BRS invariance.

On the other hand, in our formulation we are cutting thermal 
fluctuations only, which live in the on-shell sector of the theory, as 
we noticed after eq. (\ref{freeprop}). As a consequence, BRS invariance 
is preserved, and it  can be shown explicitly that the extra 
contributions to the ST identities vanish identically~\cite{noi2}.

From a computational point of view, this allows us to use BRS invariance 
as a powerful constraint in approximating the exact evolution equations.
From a physical point of view, we have a tool to derive an effective, 
gauge invariant, field theory, even for a non-zero value of the cut-off.

\section{Application: thermal masses in SU(N)}
In this section I will illustrate some preliminar results on the gluon
self-energy in SU(N)~\cite{Denis}. 
In particular I will concentrate on the 
longitudinal (or Debye) and transverse (or magnetic) masses, defined as
\beqra
m_L^2 &=&\Pi_L(q_0=0, |\vec{q}|^2=-m_L^2)\,,\nonumber\\
m_T^2 &=&\Pi_T(q_0=0, |\vec{q}|^2=-m_T^2)\,,
\label{masse}
\eeqra
where $\Pi_{L,T}$ are obtained from the self-energy $\Pi^{\mu \nu}$ as
$\Pi_L=\Pi^{00}$ and $\Pi_T=-1/2\, \Pi^{ii}$.
The definition (\ref{masse}) is gauge-independent, as shown in 
ref.~\cite{KKR}. Since the cut-off does not break BRS invariance, the
 same is true even for $\L\neq 0$.
In order to preserve this property, approximations to the full propagator
and vertices appearing in the exact RG equations must respect ST 
identities to the required  accuracy.
In the spirit of using the RG to construct an effective theory valid at sclaes
 larger than $1/gT$, we will employ a `minimal approximation scheme'  in which:
{\it i)} we use `HTL inspired' propagator and vertices, and 
{\it ii)} we rotate to imaginary time and modify the zero mode only, using 
tree-level quantities for the other ones.
In practice, we approximate the flow equation, which has the structure
\beq
\L\frac{\partial\:\:\:\:\:}{\partial\:\L} \Pi^{\mu\nu} = - 
i \int\frac{dk_0}{2\pi} N(k_0) {\rm Disc} \; F^{\mu\nu}(k_0)\;,
\eeq
with
\beq
T \sum_n F^{\mu\nu}_0 (z=2i\pi n T) - \int\frac{dk_0}{2\pi} F^{\mu\nu}_0(k_0)
+T\left[F^{\mu\nu}_{HTL}(z=0) - F^{\mu\nu}_{0}(z=0)\right]\;,
\eeq
where in $F^{\mu\nu}_{HTL}$ HTL-inspired proagator and vertices have been 
used, whereas in $F^{\mu\nu}_0$ they appear at the tree level.

The `HTL-inspired' propagator we need is
\beqra
&&\left.\Delta_{\mu\nu}\right|_{k_0=0} =
\frac{1}{|\vec{k}|^2+m_{L,\L}^2} 
g_{\mu 0} g_{\nu 0} \nonumber\\
&& \left.+ \frac{1}{|\vec{k}|^2+m_{T,\L}^2} \left(g_{\mu\nu} - 
g_{\mu 0} g_{\nu 0} + \frac{k_\mu k_\nu}{|\vec{k}|^2}\right) 
-\alpha \frac{k_\mu k_\nu}{|\vec{k}|^2}\right|_{k_0=0}\,,
\eeqra
where, compared to the true HTL propagator we have $\L$-dependent 
$m_L$ and $m_T$
(from now on we will omit the $\L$-dependence of the various variables).

The tree-level trilinear vertex does not satisfy the ST identity if 
$m_T\neq 0$. We correct it according to  eq. (12) of 
ref.~\cite{Nair}. Moreover,
it is easy to realize that, due to the breaking of Lorentz invariance 
in the
thermal bath, there is a different running for the vertex with all  the three
 space-like indices and those with at least one time-like index. Accordingly,
 we introduce two different couplings, $g_L$ and $g_T$.

Now we have a system of differential equations for $m_L$, $m_T$, $g_L$, $g_T$,
and the wave function renormalizations. In order to make contact with 
the results of perturbation theory and HTL, we first consider the 
running of the masses only.
In the picture on the left we plot $m_L/T$ vs. $\L/T$. 
The coupling constants have been 
kept fixed to $g=0.8$. 
For $\L/T \gg 1$ there is no renormalization, since  the thermal
modes coming in equilibrium are Boltzmann suppressed. As $\L/T \sim 1$ the 
HTL contribution is quickly built up (at leading order, 
$m^{HTL}_L/T= g$ in SU(3)), and if we 
stop the running at $\L/T = O(g)$ we recover the result of the 
HTL effective field theory.
Below this region the crucial question is the value of $m_T$. In perturbation
theory, a $m_T=O(g^2 T)$ is usually invoked as a infrared regulator. It 
corresponds to the lowest line, which exhibits decoupling 
for $\L < m_T$ and leads to a finite result in good agreement 
with the HTL one.
On the other hand, if $m_T=0$, $m_L$ increases logarithmically in
the infrared  and  the HTL result becomes completely unreliable at 
lower scales. 
The dashed line corresponds to the result obtained by coupling
the RG equation for $m_T$ to that for $m_L$. Assuming 
$m_T(T=0) = O(\L_{QCD})$ a $m_T\ll g^2 T$ is generally found at finite 
temperature (for the parameters used in the plot we get
$m_T/g^2T=5.3\cdot 10^{-2}$). As a consequence, the decoupling 
takes place `later'  and the finite result for $m_L$ is sizably higher
than the HTL one.

The running of the couplings is shown in the picture on the right.
Notice the impressive difference in the thermal effects on $g_L$ and 
$g_T$. This is due to the fact that graphs with only transverse gluons 
in the loop contribute to the running of $g_T$ but not to that 
of $g_L$.
The non-vanishing $m_T$ leads to  decoupling in the infrared, 
but it takes place when the couplings have already reached highly
non-perturbative values. 

The failure of HTL perturbation theory in describing the long-distance 
physics in a non-abelian plasma is  explicit in Fig.~1. The RG in the 
real time formalism   
provides a suitable framework to derive the effective theory for the 
long wavelength, non-perturbative modes.

\begin{figure}
\begin{center}
\mbox{\epsfig{file=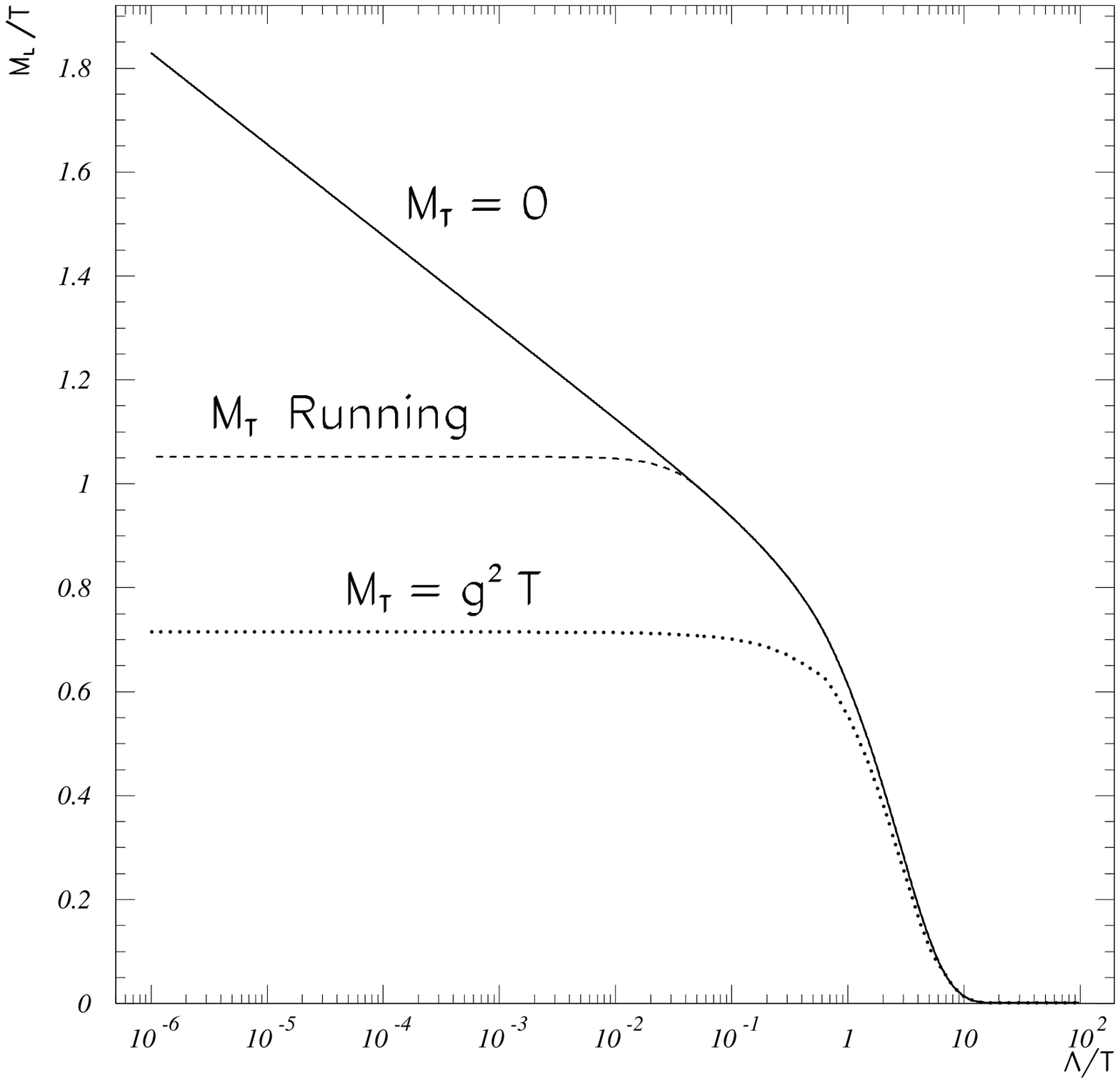,width=2.1in}}
\hspace{0.2in}
\mbox{\psfig{file=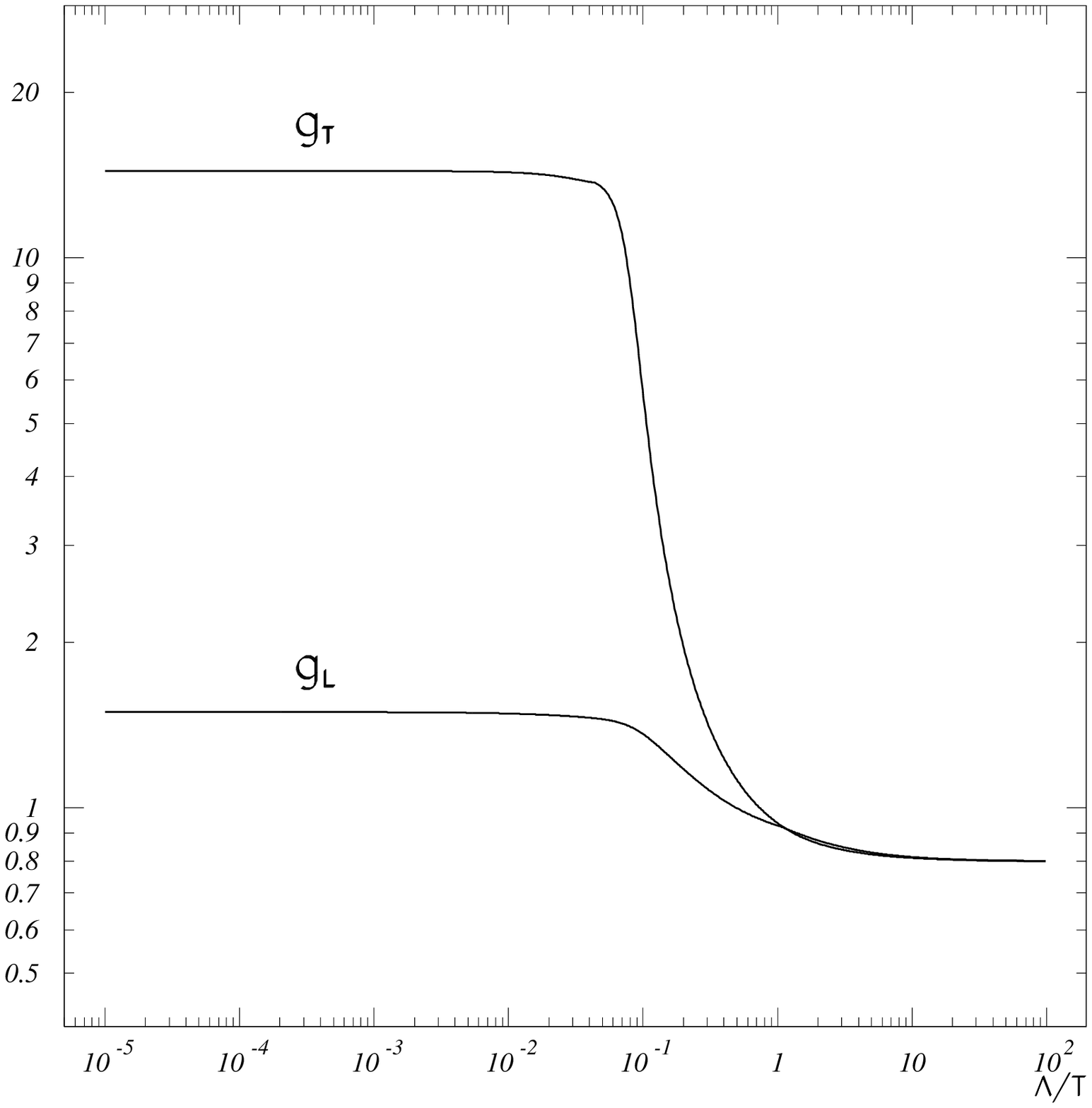,width=2.1in}}
\end{center}
%\hbox{\epsfig{file=glgt08log.ps,width=2.1in}}}
\caption{On the left: running of $m_L/T$ vs. $\L/T$ for $m_T=0$ (solid 
line), $m_T$ running (dashed line), and $m_T= g^2 T$ (dashed 
line).
On the right: running of the couplings $g_L$, and $g_T$ (see text) with
$m_L$ and $m_T$ fixed at the values obtained from the  RG.}
\end{figure}
%\begin{figure}
%\begin{center}
%\end{center}
%\caption{.}
%\end{figure}


\section*{References}
\begin{thebibliography}{99}
\bibitem{noi1}
M. D'Attanasio and M. Pietroni, {\em Nucl. Phys. B\/} {\bf 472} (1996) 711.
\bibitem{noi2}
M. D'Attanasio and M. Pietroni, hep-th/9611038 to appear on 
{\em Nucl. Phys. B}.
\bibitem{RT}
N.P. Landsman and Ch.G. van Weert, {\em Phys. Rep.\/} {\bf 145} (1987) 141.
\bibitem{Pol} 
J. Polchinski, {\em Nucl. Phys. B\/} {\bf 231} (1984) 269; 
M. Bonini, M. D'Attanasio and G. Marchesini, {\em Nucl. Phys. B\/}
{\bf 409} (1993) 441;
\bibitem{BRS}
M. Bonini, M. D'Attanasio and G. Marchesini, {\em Phys. Lett. B\/}
{\bf 346} (1995) 87; {\em Nucl. Phys. B\/} {\bf 437} (1995) 163;
U. Ellwanger, {\em Phys. Lett. B\/} {\bf 335} (1994) 364;
M. Reuter and C. Wetterich {\em Nucl. Phys. B\/} {\bf 401} (1993) 567.
\bibitem{Denis} D. Comelli and M. Pietroni, in preparation.
\bibitem{KKR}  R. Kobes, G. Kunstatter and A. Rebhan,  {\em Phys. Rev. Lett.} 
{\bf 64} (1990) 2992; {\em Nucl. Phys. B\/} {\bf 355} (1991) 1.
\bibitem{Nair} G. Alexanian and V.P. Nair,  
{\em Phys. Lett. B\/} {\bf 352} (1995) 439.


\end{thebibliography}
\end{document}